\documentclass[preprintnumbers,amsmath,amssymb]{revtex4}
\usepackage{epsfig}
\usepackage{bm}

\newcommand{\vph}{\varphi}

\begin{document}

\title{An efficient algorithm simulating a macroscopic system at the critical point}

\author{N.~G.~Antoniou}
\author{F.~K.~Diakonos}
\author{E.~N.~Saridakis}
\email{msaridak@phys.uoa.gr}
\author{G.~A.~Tsolias}
\affiliation{Department of Physics, University of Athens, GR-15771 Athens,
Greece}

\date{\today}

\begin{abstract}
It is well known that conventional simulation algorithms are
inefficient for the statistical description of macroscopic systems
exactly at the critical point due to the divergence of the
corresponding relaxation time (critical slowing down). On the
other hand the dynamics in the order parameter space is simplified
significantly in this case due to the onset of self-similarity in
the associated fluctuation patterns. As a consequence the
effective action at the critical point obtains a very simple form.
In the present work we show that this simplified action can be
used in order to simulate efficiently the statistical properties
of a macroscopic system exactly at the critical point. Using the
proposed algorithm we generate an ensemble of configurations
resembling the characteristic fractal geometry of the critical
system related to the self-similar order parameter fluctuations.
As an example we simulate the one-component real scalar field
theory at the transition point $T=T_c$ as a representative system
belonging to the $3-D$ Ising universality class.
\end{abstract}

\maketitle

\section{Introduction}
The traditional simulation algorithms of statistical mechanics are
proven to be inefficient for the simulation of the statistical
properties of macroscopic systems at a critical point. This is a
consequence of the well-known mechanism of critical slowing down
leading to a divergence of the relaxation time, independently of
the algorithm used, when the critical point is reached
\cite{Barkema99}. Therefore, the generation of an ensemble of
configurations carrying the properties of the critical system is
in practice a very difficult task as the onset of equilibrium is
prerequisite. On the other hand, due to the universal character of
the correlations at the critical point, the dynamics of the order
parameter, described effectively through the averaged action
 \cite{Wetter93} or the constrained effective potential
 \cite{Kyriak},
 acquire a very simple form \cite{Tetrad02,Tsypin94} reflecting the
 self-similarity of the associated fluctuation pattern. The
  consequence of self-similarity at the macroscopic level is at
   best reflected in the fractal geometry of the domains with
   constant magnetization in an Ising ferromagnet.
 However, fractality in the strict mathematical sense is only defined
 for an ideal critical system embedded in a continuous space. In real physical systems
 fractal geometry occurs only partially between some well defined scales.
  Therefore, physical fractals, contrary to the corresponding exact
  mathematical sets, can be defined also on equidistant lattices
 facilitating their realization through
numerical simulations. On the other hand, a realistic algorithm
generating critical configurations of a system at its transition
point should also reproduce the (partial) fractal geometry of the
critical clusters. There are several ways to produce sets with
prescribed fractal dimension found in the literature
\cite{Mandel83}. The associated algorithms can either be of
stochastic \cite{Alemany94} or deterministic nature \cite{Cantor}.
However, although fractal geometry is strongly related to critical
phenomena, there is, up to now, no algorithm capable to link
directly this geometrical property with the statistical mechanics
of a critical system.

The present work attempts a step in this direction. The main goal
is to develop an efficient algorithm for the simulation of a
macroscopic system at the critical point. In fact we propose a
method able to generate an ensemble of microstates of the
considered system carrying its basic critical properties such as
self-similar order parameter fluctuations and algebraically
decaying spatial correlations \cite{Stanley}. Using the
representation of the partition function in terms of the effective
free energy at the critical point we proceed within the saddle
point approximation. Then we show that the associated functional
measure can be expressed as a summation over piecewise constant
configurations of the order parameter within domains (clusters) of
variable size covering the entire critical system. We use these
configurations to calculate the ensemble averaged density-density
correlation function leading to a power-law form characteristic of
a fractal set. Thus we recover a relation between the fractal
geometry of the critical clusters and the canonical ensemble of
critical configurations describing the statistical properties of
the system at the transition point.

The generation of an ensemble of critical configurations is of
relevance for the study of the evolution of a critical system
under the influence of an external potential removing it from the
initial critical state. In this case the determined critical
ensemble is introduced as a set of initial conditions for the
corresponding dynamical problem \cite{manosarchive}. Besides, one
can also consider the time evolution of a critical system under
the constraints of thermodynamical equilibrium so that the
property of self-similarity sustains for an infinitely long time
interval. This situation could be actually realized in the case of
biological systems for which there exist several indications
 that they are permanently on a critical state \cite{Chialvo06}.

The paper is organized as follows: in section II we describe the
basic ingredients of the proposed algorithm for the construction
of a random fractal measure on a lattice. In section III,
subsection (A), we study in detail the one dimensional case. For a
genuine $1-D$ system with short range interactions, critical
behavior is not possible \cite{Peierls}. However, it turns out
that the $1-D$ version of the algorithm is optimal for the
illustration of the basic steps of the procedure. In addition, an
$1-D$ effective theory describing the projection of a $3-D$
critical system, can in principle carry imprints of the critical
state. In subsection (III B), we extend the proposed algorithm for
the general $D-$dimensional case. Finally, in section IV we
summarize our concluding remarks.

\section{Saddle point approach for single cluster $\vph$-configurations at the critical point}

There are two extreme situations where the division of the whole
system to subsystems and the subsequent study of one of them, is a
good approximation. The first is when the interactions are very
weak so that statistical independence is valid, and the whole
system can be assumed to be constituted by separated building
blocks. On the opposite limit, in a critical system, where the
correlations are very strong, we expect the emergence of
self-similarity and the formation of fractal structures. In this
case investigation of a small region offers information for the
entire system. Thus, at the critical point, it is natural to
consider the partition function $Z_{\Omega}$ of an open subdomain
$\Omega$ of a critical cluster. Assuming that the order parameter
is an one-component real field $\vph$, $Z_{\Omega}$ is given as:
\begin{equation}
Z_{\Omega}=\int {\cal{\delta}}[\vph]
e^{-\Gamma[\vph,\Omega]},\label{partfunct}
\end{equation}
in terms of the effective action:
\begin{equation}
\Gamma[\vph,\Omega]=\int_{\Omega} d^Dx \{ \frac{1}{2} (\nabla \vph(\vec{x}))^2 + g
\vph^{\delta+1}(\vec{x})  \}, \label{effact}
\end{equation}
expected to be valid at the critical point $T=T_c$
\cite{Tetrad02,Tsypin94}. In (\ref{effact}) $\delta$ is the
isothermal critical exponent \cite{Stanley}. To calculate the
partition function (\ref{partfunct}) one may use the saddle point
approach developed in \cite{Anton98} where it is shown that the
functional sum in $Z_{\Omega}$ is saturated by the summation over
the saddle point solutions of the effective action (\ref{effact}).
These solutions encompass power-law singularities. In the sum
(\ref{partfunct}) the only significant contribution comes from
those saddle points for which the corresponding singularities lie
outside the region $\Omega$. In fact, if the distance of the
location of the singularities from the boundary of the domain
$\Omega$ increases, the corresponding statistical weight increases
too. In this case the saddle point solutions can be well
approximated by constant functions inside $\Omega$ \cite{Anton98}.

Since in the present work we are mainly interested in the
calculation of the density-density correlation function, it is
necessary to extend the saddle point method to the case of the
generalized effective action:
\begin{equation}
\Gamma_G[\vph,\Omega]=\int_{\Omega} d^Dx \{ \frac{1}{2} (\nabla \vph(\vec{x}))^2 + g
\vph^{\delta+1}(\vec{x})+j\vph(\vec{x})\delta(\vec{x})  \},
\label{effactjj}
\end{equation}
involving a source term at $\vec{x}=0$. The saddle point solutions of (\ref{effactjj}) obey the evolution equation:
\begin{equation}
\nabla^2 \vph(\vec{x}) -
g(\delta+1)|\vph(\vec{x})|^{\delta}=j\delta({\vec{x}}).
\label{saddle}
\end{equation}
Interpreting $\vert \vph(\vec{x}) \vert$ as density the associated
ensemble of saddle point solutions is statistically identical to
the density-density correlation function $\langle
|\vph(\vec{x})\vph(\vec{0})| \rangle$, in the limit
$j\rightarrow0$ \cite{huang}. Note that the presence of the source
at $\vec{x}=0$ does not affect the generality and we can use any
other point as reference.

In order to simplify the presentation of the generalized saddle
point calculation, in the following we will describe the
one-dimensional case. However, as mentioned above, critical
behavior, in the absence of long range interactions, occurs only
in higher dimensional systems. Thus, the $1-D$ case should be
considered as a valuable explanatory tool allowing also for
analytical results, or as an effective description of the $1-D$
projection of a higher dimensional critical system. In analogy
with the treatment in \cite{Anton98}, eq.(\ref{saddle}) is solved
as an initial value problem. The solution in one dimension is
given implicitly in terms of the function:
\begin{equation}
H(\vph,E)= \frac{2}{(\delta+1)}g^{-\frac{1}{\delta+1}}\sqrt{E+g|\vph|^{\delta+1}}E^{-\frac{\delta}{\delta+1}}
\,F\left(\frac{1}{2},\frac{\delta}{\delta+1},\frac{3}{2},1+\frac{g|\vph|^{\delta+1}}{E}\right),\label{Hf}
\end{equation}
where $E$ is a parameter depending on the assumed initial
conditions and having the interpretation of the total energy of
the system. $F\left(\alpha,\beta;\gamma;z\right)$ is the usual
hypergeometric function \cite{RZ}. For $x \geq 0$ the saddle point
solutions are the inverse function of:
\begin{equation}
x=H(\vph(x),E_{+})-H(\vph(0),E_{+}),\label{sol1}
\end{equation}
while for $x<0$ they are the inverse function of
\begin{equation}
x=H(\vph(x),E_{-})-H(\vph(0),E_{-}).\label{sol2}
\end{equation}
Due to the source term the energy of the system varies in the two
half-spaces and the corresponding energy difference is determined
by the formula:
\begin{equation}
E_{-}=E_{+}+\frac{j^2}{2}+j\sqrt{2\left(E_{+}+g|\vph(0)|^{\delta+1}\right)},
\end{equation}
In the limit $j\rightarrow0$, $E_{-}=E_{+}=E$ and these solutions simplify to those presented in \cite{Anton98}.

\begin{figure}[h]
\begin{center}
\mbox{\epsfig{figure=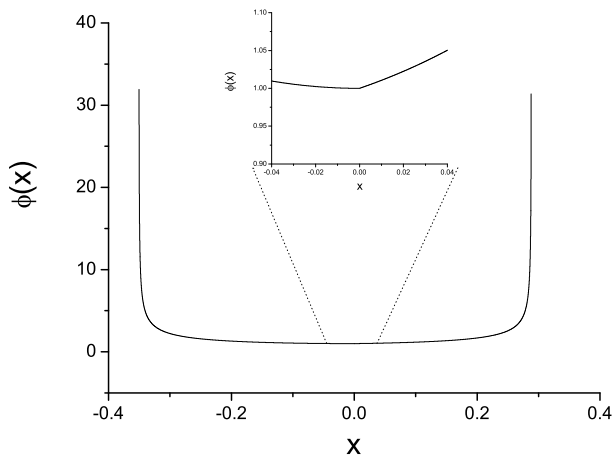,width=9cm,angle=0}}
\caption{\it A typical generalized saddle point solution, obtained
analytically, for the $1-D$ case. The discontinuity of the
derivative at $x=0$ is clearly seen in the inset figure.}
\label{analsol}
 \end{center}
 \end{figure}

Similarly to the previous discussion, a typical generalized saddle
point solution possesses two singularities which can be seen in
the graphical presentation in fig~\ref{analsol}.
Additionally it possesses a discontinuity in the field derivative
at $x=0$ which vanishes in the limit $j\rightarrow0$. In analogy
to the case without the source term, the summation over the
generalized saddle points is also dominated by configurations for
which the singularities lie beyond the range of the cluster.
Therefore, the solutions (\ref{sol1}),(\ref{sol2}) which
contribute significantly to the system's partition function can
also be
 well approximated by a constant function within the cluster.

The saddle point solutions in the higher dimensional case are
impossible to be found analytically. However numerical
calculations reveal a similar behavior, that is almost constant
configurations with singularities forming a closed $D-1$ subspace
around the location of the source, which can be handled as
described above. Therefore, the summation over the saddle point
solutions in the single cluster partition function, in any space
dimension $D$, can be well approximated by an ordinary integration
over constant field configurations.

The fractal geometry of the critical clusters is revealed through
the power-law dependence of the density-density correlation
function which in turn leads to a power-law dependence of the mean
``mass" $m(R)$ within a domain of radius $R$ around
$\vec{x}_0=\vec{0}$:
\begin{equation}
m(R)\sim R^{D_f} \label{masspower},
\end{equation}
with the exponent $D_f$ in (\ref{masspower}) being the so-called
fractal mass dimension \cite{Mandel83,Vicsek, Falconer}.
Alternatively, using the saddle-point ensemble defined above the
mean ``mass" can be calculated as:
\begin{equation}
m(R)=\langle\int_{R} |\vph(\vec{x})\vph(\vec{0})|\, d^Dx\rangle.
\label{massdimen}
\end{equation}
For a critical system at thermal equilibrium $D_f$ is related to
the critical exponent $\delta$, appearing in the effective action
(\ref{effact}), as well as the dimension $D$ of the embedding
space through \cite{stinchcombe}:
\begin{equation}
D_f=\frac{D\delta}{\delta+1} \label{fracdim}.
\end{equation}
The analysis at the level of one cluster is not sufficient for the
description of the properties of the entire critical system.
Therefore, after obtaining the set of saddle-point solutions of
the generalized action (\ref{effactjj}), it is necessary to use it
in order to generate an ensemble of configurations valid for the
global system. The systematic way to perform this extension is
explained in detail in the next section.

\section{Generation of multi-cluster configurations for the critical system}

In order to generate an ensemble of global configurations for the
critical system within the saddle point approach discussed in the
previous section one adopts the following picture: The entire
system is composed of weakly correlated clusters of different size
up to the correlation length \cite{Anton98} (which at the critical
point is expected to be of the order of the lattice size). Thus,
for given number and size of clusters, the partition function of
the system can be decomposed in the product of the partition
functions of each cluster.  The sum over the microstates in the
partition function of one cluster with given size is calculated
through the summation over the corresponding saddle points.  To
saturate the summation over microstates in the total partition
function one has to sum over the possible number of clusters as
well as the corresponding possible cluster sizes. Therefore, the
functional measure of the global partition function can be
expressed as:
\begin{equation}
\int{\cal{D}}[ \vph ]e^{-\Gamma[ \vph,\Omega_T ]}\equiv \sum_{M} \sum_{ \{\Omega_i \} } \frac{1}{M!}\prod_{i=1}^M
\int{\cal{D}}[ \vph_i ] e^{-\Gamma[\vph_i,\Omega_i]} \label{globparf},
\end{equation}
where $\cup_{i=1}^M \Omega_i = \Omega_T$ and $\Omega_i \cap
\Omega_j = \emptyset$ $\forall i,j$ with $i \neq j$, in order to
ensure that the $M$ considered clusters of volume $\Omega_i$ are
uncorrelated and cover the entire system of volume $\Omega_T$. The
functional integration $\int{\cal{D}} [ \phi_i ]$ is defined as
the sum over the saddle point solutions within the $i$-th cluster.
According to the discussion in section II the dominant
contribution to this sum comes from constant field configurations
inside each cluster. Thus:
$$\int{\cal{D}} [ \vph_i ] \equiv \int_{-\infty}^{\infty} d \vph_i.$$
In subsection (III A) we develop an algorithm for the integration
in (\ref{globparf}) based on the saddle point approach  in $1-D$
discussed so far. The algorithm is extended for higher dimensional
systems in subsection (III B). Emphasis is given in the
determination of observables carrying the fractal properties of
the critical system.

\subsection{One Dimensional Case}

Our aim is to produce an ensemble of configurations, leading to a
fractal mass dimension dictated by the power-law form of the
density-density correlation of the critical system, in realistic
computational times. According to the saddle point approach, a
suitable basis to express the configurations of the critical
system are the piecewise constant configurations, where the
domains of constant value of the field $\vph(x)$ correspond to
different clusters. In fact this coarse-graining procedure clearly
neglects any internal structure of the critical clusters. However
the entire set of these configurations, weighted with the
corresponding Boltzmann factor calculated from the effective
action, allows for the definition of suitable observables at the
level of ensemble averaged quantities, reflecting the fractal
geometry of the critical system. The main observable used in the
present work is the mean ``mass" in a domain of radius $R$,
centered at $x_0$, defined in eq.(\ref{massdimen}). Therefore, the
calculated value of the mass dimension $D_f=\frac{\ln m(R)}{\ln
R}$ will serve as a consistency check for the validity of our
procedure. To construct the multi-cluster piecewise constant
configurations we firstly perform a random partitioning of the
$N$-site lattice in a random integer number of elementary clusters
$M$ of different length. This is achieved using the Random
Partition of an Integer n (RanPar) algorithm of \cite{ranpar}.
Moreover, we use the Random Permutation of n Letters (RanPer)
algorithm from the same reference, in order to permute and
randomly select one specific partition. We point out here that
this step is one of the most time consuming of the whole
procedure, especially for $N\gtrsim4\times10^3$. In the end, we
come up with a random partitioning of the lattice into several
clusters, and each cluster consists of a different number of
lattice sites.

Continuing, we give a random constant value of the field (using a
uniform distribution in the interval $[-\vph_{max},+\vph_{max}$])
to the lattice sites within each cluster, and for this piecewise
constant configuration we calculate the $1-D$ effective action
(\ref{effact}) as:
\begin{equation}
\Gamma_{0}[\vph]=\Gamma_{00}[\vph]=\sum^N_i \alpha\, \left\{
\frac{1}{2} \left(\frac{\vph_{i+1}-\vph_{i-1}}{2\alpha}\right)^2 +
g \vph_i^{\delta+1} \right\}, \label{effactlattice}
\end{equation}
using periodic boundary conditions. Obviously, the derivative is
zero inside each cluster and becomes relevant only in their edges.
The choice of $\vph_{max}$, of the lattice spacing $\alpha$ and of
the coupling $g$, will be discussed later.

As a next step we randomly alter the field value of the first
cluster and we recalculate the effective action $\Gamma_{01}$. If
the new effective action $\Gamma_{01}$ is smaller than the initial
$\Gamma_{00}$ we accept the new field value of the first cluster.
On the other hand, if $\Gamma_{01}\geq\Gamma_{00}$ we compare
$e^{(\Gamma_{00}-\Gamma_{01})}$ with a uniformly distributed
random number $z$ in the interval $[0,1)$, and if
$e^{(\Gamma_{00}-\Gamma_{01})}>z$ we also accept the change,
otherwise we reject it. In this way the field configuration is
weighted by a factor of $e^{-\Gamma[\vph]}$. We repeat this step
changing the field value of the second cluster and calculating the
new effective action $\Gamma_{02}$, and so on, until we have
covered each one of the $M$ clusters. In the end of this
Metropolis algorithm \cite{Metropolis53} we come up with a new
field configuration and the corresponding action
$\Gamma_{1}=\Gamma_{0M}$. We repeat this procedure at least
$k_{iter}$ times in order to achieve equilibrium. The
equilibration time $k_{iter}$ is defined as the minimum number of
steps required in order to reach a stationary state, in the sense
that variations of $\vph$ for $n_s$ ($\approx O(10^3))$ successive
algorithmic steps lead to a standard deviation of the mean value
of $\vph$ and $\vph^2$ less than $0.5 \%$. The first configuration
fulfilling this criterion is actually the first member of the
ensemble of critical configurations and gets recorded.

Repetition of the whole procedure will form the complete critical
ensemble of the field configurations. Our algorithm allows for the
use of a specific lattice partition more than once, in order to
save computational time, since the partitioning procedure, using
RanPar algorithm, as well as the equilibration process are
time-consuming. Doing so, the necessary minimum iteration number
$l_{iter}$, separating in time two successive statistically
independent configurations, is much smaller than $k_{iter}$
(usually $l_{iter}\approx \frac{k_{iter}}{100}$). However, the
configuration space of the system is more efficiently covered
using different lattice partitions, and therefore an optimization
is possible leading to the use of the same partition for 10-50
times for an ensemble of $\geq10^4$ configurations.

Finally, a comment concerning the choice of the various
computational parameters has to be made. Firstly, the minimum
iteration number $l_{iter}$ can be calculated as usual in terms of
the autocorrelation function $G(m)$:
\begin{equation}
G(m)=\langle \bar{\vph}_l\bar{\vph}_{l+m}\rangle-\langle
\bar{\vph}_l\rangle\langle \bar{\vph}_{l+m}\rangle
 , \label{acf}
\end{equation}
where $\bar{\vph}_l$ is the spatial average of the field in a
configuration obtained by a single Metropolis step and time
averaging for a large number of Metropolis steps ($\gtrsim10^4$)
is performed. If $m^*$ is the characteristic decay time of $G(m)$
then we choose $l_{iter} \approx 5 m^*$.

Secondly, we have to find $\vph_{max}$, which determines the
optimal range of $\vph$ values to be used in the simulation.
Fortunately, there is a physical reason which constrains
$\vph_{max}$. Due to the potential term in the effective action
(\ref{effact}), the statistical weight of configurations involving
large $\vph$ values are super-exponentially suppressed. Thus in
thermodynamic equilibrium only configurations with field values
within a narrow interval around zero will be statistically
significant.

For every choice of $\vph_{max}$ we produce a large number
($\sim10^4$) of configurations and  from this ensemble we
calculate the average of the absolute value of the field
$\langle|\vph|\rangle_i$ for every lattice site $i$. The quantity
$\langle|\vph|\rangle_i$ versus $i$ corresponding to each
$\vph_{max}$ is almost constant (with an expected noise that
decreases increasing the ensemble population), with two peaks in
the two lattice ends. While $\vph_{max}$ increases, a shift to
larger constant $\langle|\vph|\rangle_i$ values is observed.
However, above a threshold value for $\vph_{max}$ a saturation
takes place. This behavior is expected since for small
$\vph_{max}$ statistically significant field values contributing
to the partition function are left out of the corresponding sum,
while for $\vph_{max}$'s larger than a specific value, the
additional field values have vanishingly small contribution
suppressed by the weight $e^{-\Gamma[\vph]}$. So $\vph_{max}$ can
be strictly determined using a lower limit in the variation of the
$\langle|\vph|\rangle_i$ vs $i$ dependence. Knowing $\vph_{max}$
reduces substantially the computational time as the requisite
$l_{iter}$ increases rapidly with increasing $\vph_{max}$.

Lastly, we have to fix the coupling $g$ and the lattice spacing
$\alpha$. The former can be determined by considering the $1-D$
effective action as the projection of (\ref{effact}) in one
dimension, since in a genuine $1-D$ system there is no critical
point in the absence of long range interactions \cite{Peierls}.
However, technically, an ensemble of configurations reproducing
equivalently the density-density correlation of a fractal set with
prescribed fractal mass dimension ($D_f < 1$) can be constructed
although having no direct physical interpretation. Concerning the
lattice spacing $\alpha$ we have considered the thermodynamical
and the continuum limit investigating the behavior of the term
$\langle\frac{\delta\vph}{\alpha}\rangle$ for $N\rightarrow\infty$
and $\alpha=\text{const}$ or $a\rightarrow0$. In both cases we
have observed a smooth behavior, attributed to the used basis of
piecewise constant configurations, as well as to the the effect of
the statistical weight  $e^{-\Gamma[\vph]}$ of each configuration
which favors smoother ones as $N\rightarrow\infty$. Therefore, the
proposed algorithm is in fact independent of $\alpha$.  However,
if one desires to evolve the produced configurations in time, it
is necessary to use the same lattice spacing in the numerical
evolution, in order to achieve self-consistency
\cite{manosarchive}. Finally, we stress that in
 our presented results the calculated observables are expressed in units
of $\alpha$.

With the demonstrated procedure we acquire an ensemble of field
configurations generating a random fractal measure on the lattice
as a statistical property after ensemble averaging. This fractal
property (\ref{masspower}) is not directly reflected in the
geometry of their average $\langle\vph(x)\rangle$, but is produced
only through the entire ensemble. The fractal mass dimension is
determined by the power-law behavior of
\begin{equation}
m(R)=\langle\int_{R} |\vph(x)\vph(0)|\, dx\rangle\sim R^{D_f},
\label{massdimen1d}
\end{equation}
around $x_0=0$, averaged inside clusters of size $R$. The
$\langle\int_{R} |\vph(x)\vph(0)|\, dx\rangle$ versus $R$ figure
is drawn as follows: For a $x_0=0$ of a specific configuration we
find the size $R$ of the cluster in which it belongs and we
calculate the integral $\int_{R} |\vph(x)\vph(0)|\, dx$, thus
acquiring one point in the $\langle\int_{R} |\vph(x)\vph(0)|\,
dx\rangle$ vs $R$ figure. For the same $x_0=0$ we repeat this
procedure until we cover the whole ensemble, and the
aforementioned figure is formed. Taking a different reference
point $x_0$ obviously does not alter the results, since due to
translation invariance in the averaged quantities $m(x_0,R)\approx
m(x_0+l,R)$, with $l$ spanning the entire lattice. Eventually, our
algorithm generates an ensemble of field configurations with
fractal mass dimension
\begin{equation}
D_f=\frac{\delta}{\delta+1} \label{fracdim1d}.
\end{equation}

\begin{figure}[h]
\begin{center}
\mbox{\epsfig{figure=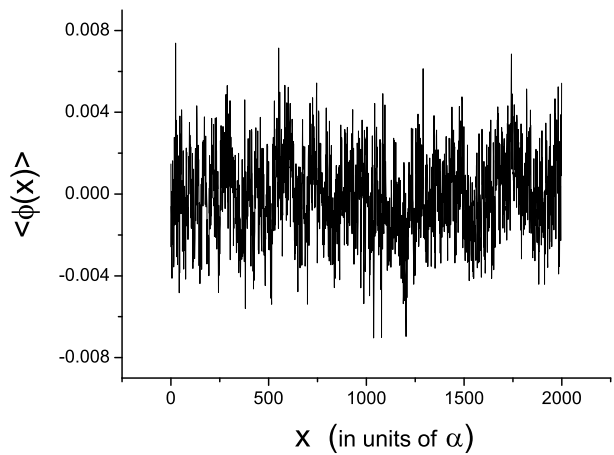,width=9cm,angle=0}}
\caption{\it The $\vph$-field on the $1-D$ lattice averaged over
the ensemble of the initial configurations.} \label{sigma0}
\end{center}
\end{figure}

\begin{figure}[h]
\begin{center}
\mbox{\epsfig{figure=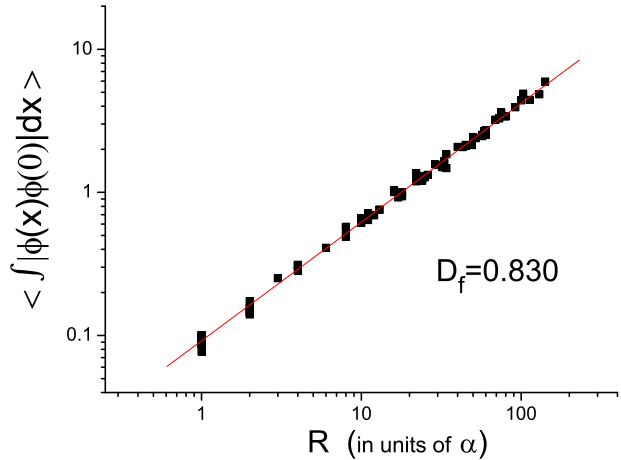,width=9cm,angle=0}} \caption{\it
$\langle\int_{R} |\vph(x)\vph(0)|\, dx\rangle$ versus $R$ for the
ensemble of $\vph$-field configurations. The slope, i.e the
fractal mass dimension $D_f$, is equal to $5/6$ within an error of
less than 0.3\%.} \label{sigma01d}
\end{center}
\end{figure}

As an application we produce an ensemble of $10^4$ one dimensional
field configurations on a $N=2000$ lattice, using $\delta=5$ and
$g=2$, in which case the theoretical value of the fractal mass
dimension according to (\ref{fracdim1d}) is $5/6$. In
fig.~\ref{sigma0}, we depict the ensemble average
$\langle\vph(x)\rangle$
having a noisy profile \footnote{Note that the fractal mass
dimension in eq.(\ref{fracdim1d}) must not be confused with the
fractal dimension of the corresponding curve, which in this case
is greater than 1.}. However, in the log-log plot of
$\langle\int_{R} |\vph(x)\vph(0)|\, dx\rangle$ versus $R$ depicted
in fig.~\ref{sigma01d},
the slope, i.e the fractal mass dimension $D_f$ according to (\ref{massdimen1d}),
is equal to $5/6$ within an error of less than 0.3\%.

\subsection{Higher Dimensional Case}

 The higher dimensional generalization of our algorithm is straightforward, preserving
 the improved mean field  approach using piecewise constant configurations. For $D>1$ we use
  configurations consisting of $D-$dimensional boxes
   as basis and the ensemble production is reduced to the Cartesian
 product of the one dimensional case. Finally, the decisive test about the proper configuration production
  will be the calculation of the fractal mass dimension (\ref{massdimen}), as in the $1-D$ case. In particular
  the $3-D$ case is physically very interesting since the effective action
(\ref{effact}) describes the order parameter dynamics of the Ising universality class at the spontaneous
 magnetization transition point. Thus, contrary to the one dimensional case, the investigated $3-D$
 system has a physical impact describing the {\it standard model} of a very common critical state.

We firstly perform a random partitioning of the $D-$dimensional
lattice in a random integer number of elementary box-shaped
clusters of different volume, each one consisting of several
lattice points. This is succeeded by applying the $1-D$
partitioning algorithm $D$ times and then taking the Cartesian
product of the $1-D$ partitions. In this case the linear size of
the lattice is reduced, leading to a significant decrease of the
partitioning algorithm computational time. For example, the time
needed for the partitioning of a $20 \times 20 \times 20$ lattice
is two orders of magnitude smaller than that of a  $2000-$site
linear lattice.

Similarly to the $1-D$ case, we assign a random constant value of
the field (in the interval $[-\vph_{max},+\vph_{max}$]) to the
lattice sites within each cluster, and for this piecewise constant
configuration we calculate the $D$-dimensional effective action
(\ref{effact}), using the corresponding generalized formula of eq.
(\ref{effactlattice}). The sum extends to every lattice site using
periodic boundary conditions, and the derivative is calculated
using the straightforward $D$-dimensional generalization. The
value of the coupling $g$ in the $3-D$ Ising effective action has
been determined in the literature ($g \approx 2$
\cite{Tetrad02,Tsypin94})  and the parameters $\vph_{max}$,
$k_{iter}$ and $l_{iter}$ are determined following the
corresponding $1-D$ steps. The only complication enters in the
specification of $\vph_{max}$, which  requires the construction of
the plot $\langle|\vph|\rangle_i$ versus lattice site $i$, as we
have mentioned in the previous subsection, becoming
computationally much more demanding in this higher dimensional
case.

As a next step, we randomly change the field value of the first
box-shaped cluster, we recalculate the effective action and using
the same criteria as in the $1-D$ case  either we accept or reject
the new field value of the first cluster. We perform these steps
until we have covered the whole lattice, and we iterate this
procedure $k_{iter}$ times, in the end of which we record one
field configuration. Repetition of the above algorithm will form
the whole ensemble of the field configurations. Finally, our
comment in the $1-D$ treatment about the multiple use of a
specific lattice partition, holds in the present case, too.
However, due to the significantly smaller partitioning
computational time, such a treatment is not necessary.

\begin{figure}[h]
\begin{center}
\mbox{\epsfig{figure=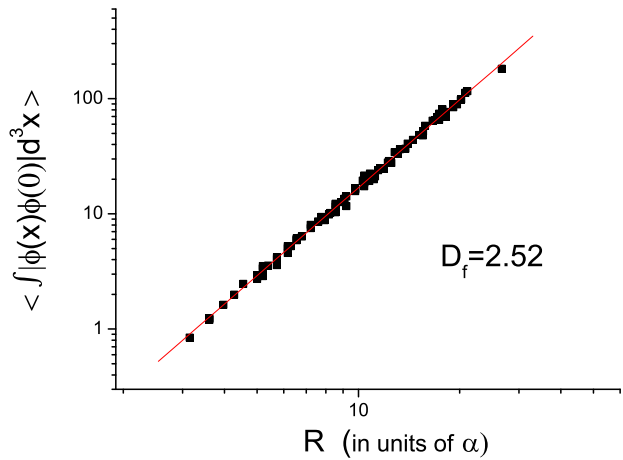,width=9cm,angle=0}} \caption{\it
$\langle\int_{R} |\vph(\vec{x})\vph(\vec{0})|\, d^3x\rangle$
versus $R$ for the ensemble of $3-D$ $\vph$-field configurations.
The slope, i.e the fractal mass dimension $D_f$, is equal to $5/2$
within an error of less than 1\%.} \label{spower0}
\end{center}
\end{figure}

This is the generalized algorithm of producing an ensemble of
$D$-dimensional field configurations generating a random fractal
measure on the lattice.  The ensemble possesses the property of
eqs. (\ref{massdimen}) and
 (\ref{masspower}), where now the fractal mass dimension is determined by
 the power-law behavior of $m(R)$ around $\vec{x}_0=\vec{0}$, averaged inside clusters
  of volume $V$. Similarly to the $1-D$ case, to acquire the
   $\langle\int_{R} |\vph(\vec{x})\vph(\vec{0})|\, d^Dx\rangle$ versus $R$
   figure we chose  $\vec{x}_0=\vec{0}$ of a
specific configuration, we find $R$ of the cluster to which it
belongs, approximated as $\propto \sqrt[D]{V}$, and we calculate
the integral $\int |\vph(\vec{x})\vph(\vec{0})|\, d^Dx$, thus
obtaining one point in the $\langle\int_{R}
|\vph(\vec{x})\vph(\vec{0})|\, d^Dx\rangle$ vs $R$ figure.
Repetition of this procedure for the whole configuration ensemble
provides all the points in Fig.~\ref{spower0}.

As an application we produce an ensemble of $10^4$ 3-dimensional
field configurations on a $20\times20\times20$ lattice, using
$\delta=5$ and $g=2$ \cite{Tetrad02,Tsypin94}. As already
discussed, this choice has a physical correspondence, since for
dimensionality $D=3$, isothermal critical exponent $\delta=5$ and
coupling $g=2$, the free energy (\ref{effact}) describes the
effective action of the $3-D$ Ising model at its critical point.
In this case, the theoretical value of the fractal mass dimension
according to (\ref{fracdim}) is $5/2$.

\begin{figure}[h]
\begin{center}
\mbox{\epsfig{figure=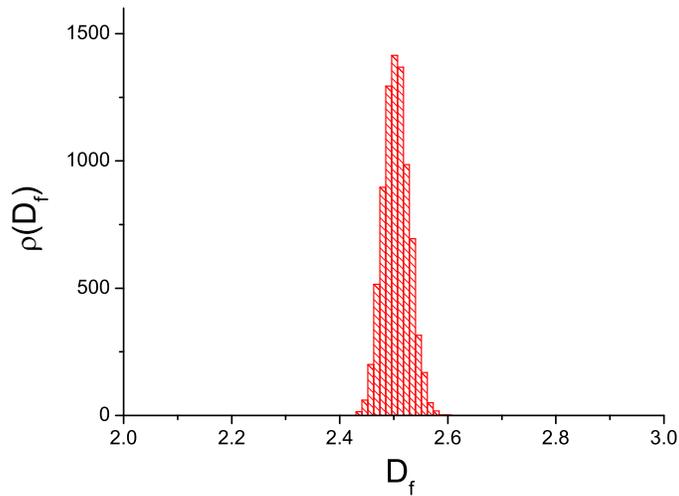,width=10cm,angle=0}}
\caption{\it $\rho(D_f)$ calculated using $8\times 10^3$ different
reference points $\vec{x}_0$ on the lattice.} \label{rhodf}
\end{center}
\end{figure}

In Fig.~\ref{spower0} we observe that in the log-log plot of
$\langle\int_{R} |\vph(\vec{x})\vph(\vec{0})|\, d^3x\rangle$ vs
$R$, the slope, i.e the fractal mass dimension $D_f$ according to
(\ref{masspower}), is equal to $5/2$, within an error of less than
1\%.
In order to test that the proposed algorithm is valid for any
lattice site, as dictated by the statistics in the considered
system, we have calculated $D_f$ using different reference points
$\vec{x}_0$ (sources in (\ref{effactjj})) on the $3-D$ lattice.
The corresponding distribution $\rho(D_f)$ is shown in
Fig.~\ref{rhodf}.
It is clearly seen that $D_f$ is almost constant with a deviation
of at most $4 \%$ for the given lattice size $N^3$, and as we have
tested, increasing $N$ this deviation decreases algebraically.

\begin{figure}[h]
\begin{center}
\mbox{\epsfig{figure=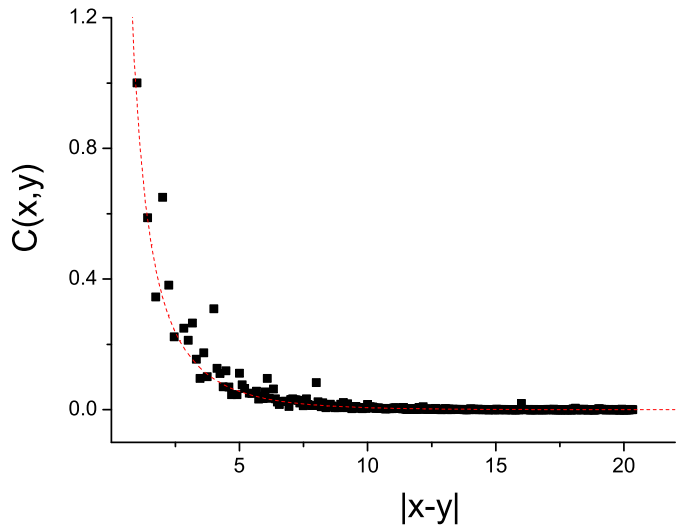,width=10cm,angle=0}}
\caption{\it $C(x,y)$ as a function of $\vert x - y \vert$
calculated using $10^4$ critical configurations for the $3-D$
one-component real scalar field (Ising). The dashed line presents
the theoretically expected result.} \label{corff}
\end{center}
\end{figure}

Finally, a last test is performed in order to check the ability of
the obtained ensemble of configurations to reproduce the
statistical properties of the critical system. Besides the
underlying fractal geometry of the critical clusters, the
two-point correlation function:
\begin{equation}
C(x,y)=\langle \vert \vph(x) \vph(y) \vert \rangle - \langle \vert \vph(x) \vert \rangle
\langle \vert \vph(y) \vert \rangle
\label{corrfun}
\end{equation}
possesses an analytically known power-law form at the critical
point. Using the constructed ensemble of configurations we have
calculated numerically the correlation function (\ref{corrfun})
and the result is shown in Fig.~\ref{corff}. The theoretical
expectation $C(x,y)\sim|x-y|^{-1-\eta}$ ($\eta\approx0.04$ for the
$3-D$ Ising universality class), corrected by a small exponential
factor incorporating finite size effects, is shown with the dashed
line. It is clearly seen that the calculated $C(x,y)$ is in very
good agreement with the analytical formula supporting further the
equivalence of the obtained ensemble of configurations with the
critical state of the considered system.

\section{Concluding remarks}

Using the saddle-point approach introduced in \cite{Anton98} we
have been able to develop an algorithm simulating the critical
state of a macroscopic system at its transition point. The method
followed here is in close analogy with the improved mean field
theory \cite{imprmeanfield} and leads to a successful and
computationally efficient description of the geometrical
characteristics of the critical clusters, defining suitable
measures to quantify this property. A particularly appealing issue
of the proposed method is that it provides a link between the
effective action at the critical point and the fractal geometry of
the formed clusters, overcoming the huge numerical effort needed
for the detailed description of fractal sets. Thus, it is the
first time, at least to our knowledge, that geometrical
characteristics as well as statistical properties of the critical
system, are incorporated in an ensemble of configurations suitable
for the study of any desired observable at the critical point. The
proposed algorithm may be of special interest for the treatment of
systems where the formed critical state is not observable but acts
as an intermediate state for the subsequent evolution of the
system. Such a scenario is likely to hold in the collision of two
heavy nuclei at high energies \cite{RW00} when the formed fireball
freezes out near the theoretically predicted QCD critical point.\\

\paragraph*{{\bf{Acknowledgements:}}} We thank N. Tetradis for useful discussions.
 One of us (E.N.S) wishes to
thank the Greek State Scholarship's Foundation (IKY) for financial
support. The authors acknowledge partial financial support through
the research programs ``Pythagoras'' of the EPEAEK II (European
Union and the Greek Ministry of Education) and ``Kapodistrias" of
the Research Committee of the University of Athens.

\end{document}